# INVISIBILITY SYSTEM USING IMAGE PROCESSING AND OPTICAL CAMOUFLAGE TECHNOLOGY

Vasireddy Srikanth[1], Pillem Ramesh[2]
[1] Student, Electronics and Communication Engineering, K L University.
[2] Asst. professor, Electronics and Communication Engineering, K L University

***Abstract:*** *Invisible persons are seen in fiction stories only, but in the real world it is proved that invisibility is possible. This paper describes the creation of invisibility with the help of technologies like Optical camouflage; Image based rendering and Retro-reflective projection. The object that needs to be made transparent or invisible is painted or covered with retro reflective material. Then a projector projects the background image on it making the masking object virtually transparent. Capturing the background image requires a video camera, which sits behind the person wearing the cloak. The video from the camera must be in a digital format so it can be sent to a computer for image processing using image based rendering technical. There are some useful applications for this simple but astonishing technology.*

***Keywords:*** Optical Camouflage Technology, Retro-Reflective material, Head Mounted Projector (HMP)

## 1. Introduction

Optical camouflage is a kind of active camouflage. This idea is very simple. If you project background image onto the masked object, you can observe the masked object just as if it were virtually transparent. The cloak that enables optical camouflage to work is made from a special material known as **retro-reflective material**. To create invisibility or transparent illusion we need a video camera, computer, projector and a combiner.

## 2. Retro-reflective material

The cloak that enables optical camouflage to work is made from a special material known as **retro-reflective material**. A retro-reflective material is covered with thousands and thousands of small beads ( Fig-1). When light strikes one of these beads, the light rays bounce back exactly in the same direction from which they came.

To understand why this is unique, look at how light reflects off of other types of surfaces. A rough surface creates a diffused reflection because the incident (incoming) light rays get scattered in many different directions.

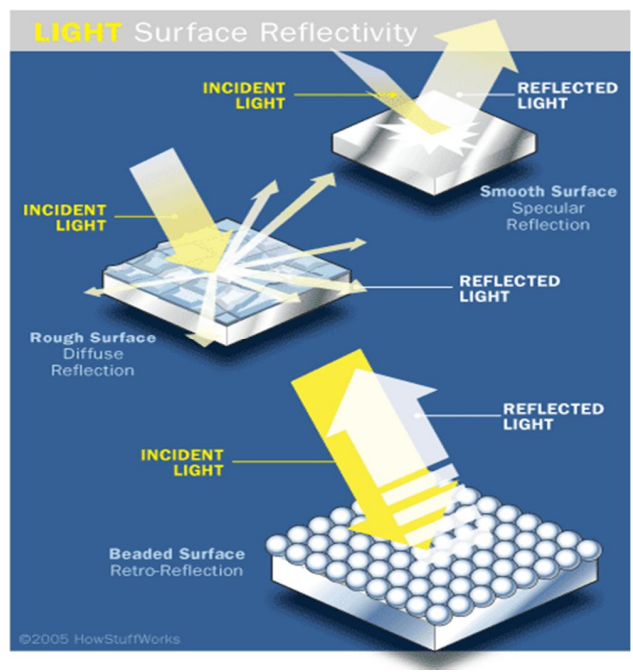

**Figure-1**

A perfectly smooth surface, like that of a mirror, creates what is known as a specular reflection -- a reflection in which incident light rays and reflected light rays form the exact same angle with the mirror surface. In retro-reflection, the glass beads act like prisms, bending the light rays by a process known as refraction. This causes the reflected light rays to travel back along the same path as the incident light rays. The result: An observer situated at the light source receives more of the reflected light and therefore sees a brighter reflection.

*Retro-reflective materials* are actually quite common. Traffic signs, road markers and bicycle reflectors all take advantage of retro-reflection to be more visible to people driving at night. Movie screens used in most modern commercial theaters also take advantage of this material because it allows for high brilliance under dark conditions. In optical camouflage, the use of retro-reflective material is critical because it can be seen from far away and outside in bright sunlight -- two requirements for the illusion of invisibility.

## 3. *The Video Camera and Computer*

### i). *Video Camera*

The retro-reflective garment doesn't actually make a person invisible -- in fact, it's perfectly opaque. What the garment does is create an illusion of invisibility by acting like a movie screen onto which an image from the background is projected. Capturing the background image requires a video camera, which sits behind the person wearing the cloak. The video from the camera must be in a digital format so it can be sent to a computer for processing.

### ii). *Computer*

All augmented-reality systems rely on powerful computers to synthesize graphics and then superimpose them on a real-world image. For optical camouflage to work, the hardware/software combo must take the captured image from the video camera, calculate the appropriate perspective to simulate reality and transform the captured image into the image that will be projected onto the retro-reflective material. This technic of image processing is called image based rendering.

## 4. *The Projector and Combiner*
### i). *The Projector*

The modified image produced by the computer must be shone onto the garment, which acts like a movie screen. A projector accomplishes this task by shining a light beam through an opening controlled by a device called an **iris diaphragm**. An iris diaphragm is made of thin, opaque plates, and turning a ring changes the diameter of the central opening. For optical camouflage to work properly, this opening must be the size of a pinhole. Why? This ensures a larger depth of field so that the screen (in this case the cloak) can be located any distance from the projector.

### ii). *The Combiner*

The system requires a special mirror to both reflect the projected image toward the cloak and to let light rays bouncing off the cloak return to the user's eye. This special mirror is called a beam splitter, or a combiner -- a half-silvered mirror that both reflects light (the silvered half) and transmits light (the transparent half). If properly positioned in front of the user's eye, the combiner allows the user to perceive both the image enhanced by the computer and light from the surrounding world. This is critical because the computer-generated image and the real-world scene must be fully integrated for the illusion of invisibility to seem realistic. The user has to look through a peephole in this mirror to see the augmented reality.

## 5. *The Complete System*

Now let's put all of these components together to see how the invisibility cloak appears to make a person transparent. The given figure-2 shows the typical arrangement of all of the various devices and pieces of equipment.

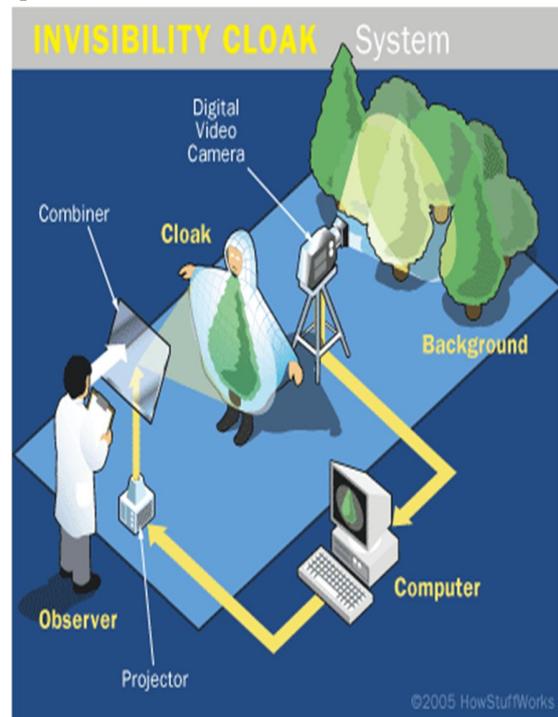

**Figure-2**

Once a person puts on the cloak made with the retro-reflective material, here's the sequence of events:

1. A digital video camera captures the scene behind the person wearing the cloak.
2. The computer processes the captured image and makes the calculations necessary to adjust the still image or video so it will look realistic when it is projected.
3. The projector receives the enhanced image from the computer and shines the image through a pinhole-sized opening onto the combiner.
4. The silvered half of the mirror, which is completely reflective, bounces the projected image toward the person wearing the cloak.
5. The cloak acts like a movie screen, reflecting light directly back to the source, which in this case is the mirror.

6. Light rays bouncing off of the cloak pass through the transparent part of the mirror and fall on the user's eyes. Remember that the light rays bouncing off of the cloak contain the image of the scene that exists behind the person wearing the cloak. The person wearing the cloak appears invisible because the background scene is being displayed onto the retro-reflective material. At the same time, light rays from the rest of the world are allowed reach the user's eye, making it seems as if an invisible person exists in an otherwise normal-looking world.

### 6. *Head-mounted Displays*

Of course, making the observer stand behind a stationary combiner is not very pragmatic -- no augmented-reality system would be of much practical use if the user had to stand in a fixed location. That's why most systems require that the user carry the computer on his or her person, either in a backpack or clipped on the Hip. It's also why most systems take advantage of head-mounted displays, or HMDs, which assemble the combiner and optics in a wearable device.

There are two types of HMDs: optical see-through displays and video see-through displays. Optical see-through displays look like high-tech goggles, sort of like the goggles Cyclops wears in the X-Men comic books and movies. These goggles provide a display and optics for each eye, so the user sees the augmented reality in stereo. **Video see-through displays**, on the other hand, use video-mixing technology to combine the image from a head-worn camera with computer-generated graphics.

In this arrangement, video of the real world is mixed with synthesized graphics and then presented on a liquid-crystal display. The great advantage of video see-through displays is that virtual objects can fully obscure real-world objects and vice versa.

The scientists who have developed optical-camouflage technology are currently perfecting a variation of a video see-through display that brings together all of the components necessary to make the invisibility cloak work. They call their apparatus a **head-mounted projector** (HMP) because the projection unit is an integral part of the helmet. Two projectors -- one for each eye -- are required to produce a stereoscopic effect.

### 7. *Real-World Applications*

While an invisibility cloak is an interesting application of optical camouflage, it's probably not the most useful one. Here are some practical ways the technology might be applied:

- Pilots landing a plane could use this technology to make cockpit floors transparent. This would enable them to see the runway and the landing gear simply by glancing down.
- Doctors performing surgery could use optical camouflage to see through their hands and instruments to the underlying tissue. See Tachi Lab: Optical Camouflage: oc-phantom.mpg to watch a video of how this might work.
- Providing a view of the outside in windowless rooms is one of the more fanciful applications of the technology, but one that might improve the psychological well-being of people in such environments.
- Drivers backing up cars could benefit one day from optical camouflage. A quick glance backward through a transparent rear hatch or tailgate would make it easy to know when to stop.

### 8. *Some images showing Invisibility*

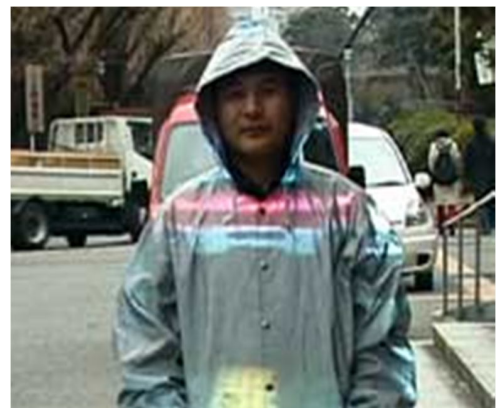

**IMAGE-1**

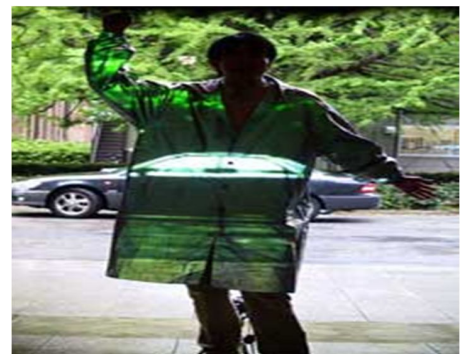

**IMAGE-2**

visible. The concept is called **Mutual Telexistence**: working and perceiving with the feeling that you are in several places at once. **Pervasive gaming** is another application where players with mobile displays move through the world while sensors capture information about their environment, including their location. This information is used to deliver users a gaming experience that changes according to where they are and what they are doing.

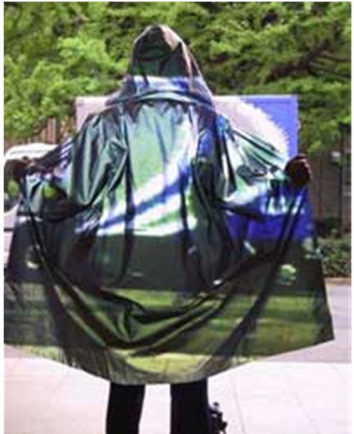

**IMAGE-3**

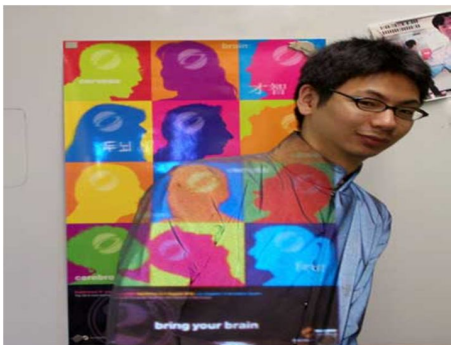

**IMAGE-4**

## 9. Conclusion

This amazing technology creates objects or human beings invisible or transparent. Though it has some limitations, it won't be for long as scientists continue to push the boundaries of the technology. One of the most promising applications of this technology, however, has less to do with making objects invisible and more about making them